\documentclass[twoside,twocolumn,english,aps, prl]{revtex4}
\usepackage[T1]{fontenc}
\usepackage[latin9]{inputenc}
\usepackage{amsmath}
\usepackage{amssymb}
\usepackage{graphicx}

\makeatletter
\@ifundefined{textcolor}{}
{%
 \definecolor{BLACK}{gray}{0}
 \definecolor{WHITE}{gray}{1}
 \definecolor{RED}{rgb}{1,0,0}
 \definecolor{GREEN}{rgb}{0,1,0}
 \definecolor{BLUE}{rgb}{0,0,1}
 \definecolor{CYAN}{cmyk}{1,0,0,0}
 \definecolor{MAGENTA}{cmyk}{0,1,0,0}
 \definecolor{YELLOW}{cmyk}{0,0,1,0}
}

\usepackage{color}
\usepackage{graphicx}

\makeatother

\usepackage{babel}
\begin{document}
\title{From Edge State Physics to Entanglement Spectrum: Studying Interactions
and Impurities in Two-Dimensional Topological Insulators\textquotedbl}
\author{Marcela Derli}
\author{E. Novais}
\email{eduardo.novais@ufabc.edu.br}

\affiliation{Centro de Ciências Naturais e Humanas, Federal University of ABC,
Brazil.}
\date{\today{}}
\begin{abstract}
We present a novel theoretical approach to incorporate electronic
interactions in the study of two-dimensional topological insulators.
By exploiting the correspondence between edge state physics and entanglement
spectrum in gapped topological systems, we deconstruct the system
into one-dimensional channels. This framework enables a simple and
elegant inclusion of fermionic interactions into the discussion of
topological insulators. We apply this approach to the Kane-Mele model
with interactions and magnetic impurities.
\end{abstract}
\maketitle
\emph{Introduction}: Two dimensional systems have captivated physicists
with their intriguing and often counter intuitive properties. Frequently,
they demand the development of new theoretical and numerical tools
for investigation. A notable historical example of this behavior is
observed in the quantum Hall effect. While the integer quantum Hall
effect can be comprehensively explained within the framework of non-interacting
particles filling Landau bands, the discovery of non-integer Hall
plateaus due to electronic interactions and disorder was a surprising
revelation. Recently the better understanding of topological properties
in the band structure of two dimensional systems have yielded new
and exciting theoretical and experimental results. However, the interplay
of topology and correlations is still a relatively open area of investigation.

In the study of strongly correlated two dimensional systems, one promising
approach is to leverage insights gained from one-dimensional models
by suitably identifying one dimensional structures in the higher-dimensional
system. Several methods can be employed to achieve this dimensional
reduction. One intuitively appealing approach is to first consider
quantum wires and subsequently introduce tunneling between them\citep{kane_fractional_2002,teo_luttinger_2014,tam_nondiagonal_2021}.
The connection to the two dimensional system is obtained using renormalization
group arguments to the tunneling amplitude. Another approach is to
deconstruct the topological system into one dimensional wires in order
to classify the topological phases\citep{neupert_wire_2014,iadecola_wire_2016}.
In a complementary view to this anisotropic starting point, we propose
a novel approach that is particularly well-suited for studying impurities
in two-dimensional topological systems.

The manuscript revolves around a pivotal concept: the spectrum of
entanglement \citep{peschel_calculation_2003} in gapped systems provides
a valuable insights into the dynamics of local operators\citep{qi_general_2012}.
This fresh perspective introduces a fictitious boundary, enabling
us to represent the original wave function as a superposition of \emph{bulk
states} and \emph{boundary states}. When focusing on the behavior
of local operators solely defined on the \emph{boundary}, it becomes
evident that only the boundary states interact with these operators.
However, this introduction of the \emph{boundary} comes at the cost
of working within a formalism at a finite fictitious temperature,
which is determined by the problem's characteristics, such as the
presence of the spectral gap.

By starting with a two-dimensional problem and transitioning to a
one-dimensional model at a finite temperature, we gain the ability
to incorporate interactions into the problem on much better footing
than in a straightforward approach within the original two-dimensional
model. This avenue of investigation presents a novel approach to studying
electronic systems, enabling a deeper comprehension of their dynamics
and properties. Furthermore, it presents a clear pathway towards two-dimensional
bosonzation, and possibly new numerical techniques, on these systems.

Though out this manuscript we consider natural unit, $\hbar=c=k_{B}=1$.
The manuscript is organized as follows: first the general formalism
is introduced. Subsequently, we delve into the analyses of the prototypical
topological band insulators, the \emph{Kane-Mele model}. We add to
this model local degrees of freedoms and electronic interactions to
investigate the interplay of magnetism, interactions and topology.
We finish with a conclusion and some possible future perspectives.

\emph{The physics of a hard-wall boundary:} In a D-dimensional system,
a hard-wall boundary is a D-1 surface that prohibits the flow of particles
and information. These properties are implemented by specific boundary
conditions on the Schrödinger equation that ensures that the current
through the boundary is zero. For instance, let's consider \emph{non-relativistic}
fermions in two dimensions with a hard boundary located at $x=0$.
To ensure that the single particle wave function $\psi\left(x,y\right)=X\left(x\right)Y\left(y\right)$
follows the hard wall conditions, we require that

\begin{equation}
\frac{\partial\psi\left(0,y\right)}{\partial x}=0,\text{or}\,\psi\left(0,y\right)=0.
\end{equation}
It was shown that for the case of \emph{relativistic} fermions\citep{lieb_two_1961,edward_witten_three_2016,fukui_theory_2020},
the \emph{no-current} condition leads to two very distinct type of
$X\left(x\right)$ functions: \emph{bulk} and \emph{boundary} solutions.
Bulk solutions correspond to standing waves and must be zero at the
boundary, 

\begin{align}
X\left(x\right) & =x_{0}\sin\left[\varepsilon x\right],
\end{align}
with $x_{0}$ and $\varepsilon$ constants. Conversely, zero energy
boundary solutions can exist. They correspond to exponentially decaying
functions

\begin{align}
X\left(x\right) & =x_{0}e^{-\frac{x}{\xi}},
\end{align}
where $\xi$ is the characteristic length of the state. The behavior
of $Y\left(y\right)$ can depend on the details of the microscopic
theory. A well known example is graphene\citep{lado_edge_2015}. In
graphene, an \emph{armchair} edge in graphene gives rise to boundary
states with energies away from the Fermi level, while a \emph{zigzag}
edge introduces states at the Fermi level with no dispersion. The
fact that we can classify solutions in the presence of a real hard
wall boundary as \emph{bulk} states and \emph{boundary} states is
a powerful insight into the spectrum of entanglement.

\emph{Spectrum of entanglement and edge states}: The concept of the
entanglement spectrum in topological systems was introduced by Li
and Haldane in 2008 \citep{li_entanglement_2008}. They investigated
a Hall system on the surface of a sphere and employed the Schmidt
decomposition to to recast the problem in terms of wave functions
defined on each hemisphere of the sphere. The logarithm of the Schmidt
coefficients was then coined the \emph{entanglement spectrum}. It
encodes the physical information on how the original wave function
is weaved at the fictitious boundary created by the choice of basis.
Remarkably, through numerical analysis, they observed that the entanglement
spectrum accurately reproduces the low energy spectrum of a genuine
boundary in a Hall bar.

Following their pioneering work, subsequent contributions have further
support for these findings\citep{casini_entanglement_2009,giudici_entanglement_2018,vidal_entanglement_2003,fidkowski_entanglement_2010,qi_general_2012,swingle_geometric_2012,chen_quantum_2013,ejima_spectral_2014,buhr_geometrical_2011}.
These studies consistently demonstrate that in topological systems
the entanglement spectrum faithfully reproduces the spectrum of genuine
edge states, albeit at finite fictitious \emph{temperature of entanglement},
$T_{E}$, and low energies.

In what follows, we assume that the initial wave function that we
aim to describe , denoted as$\left|g\right\rangle $, is the ground
state wave function of a gapped local Hamiltonian of a topological
insulator. We proceed by dividing the system into two partitions,
$\left|A\right\rangle $ and $\left|B\right\rangle $, with an arbitrary
cut. The Hamiltonian can be written as the sum of Hamiltonians pertaining
to the individual partitions, namely $H_{A/B}$ , along with the inter-partition
Hamiltonian $H_{AB}$, 
\begin{equation}
H_{0}=H_{A}+H_{B}+H_{AB}.\label{eq:initial_hamiltonian}
\end{equation}

\begin{figure}
\includegraphics[width=0.8\columnwidth]{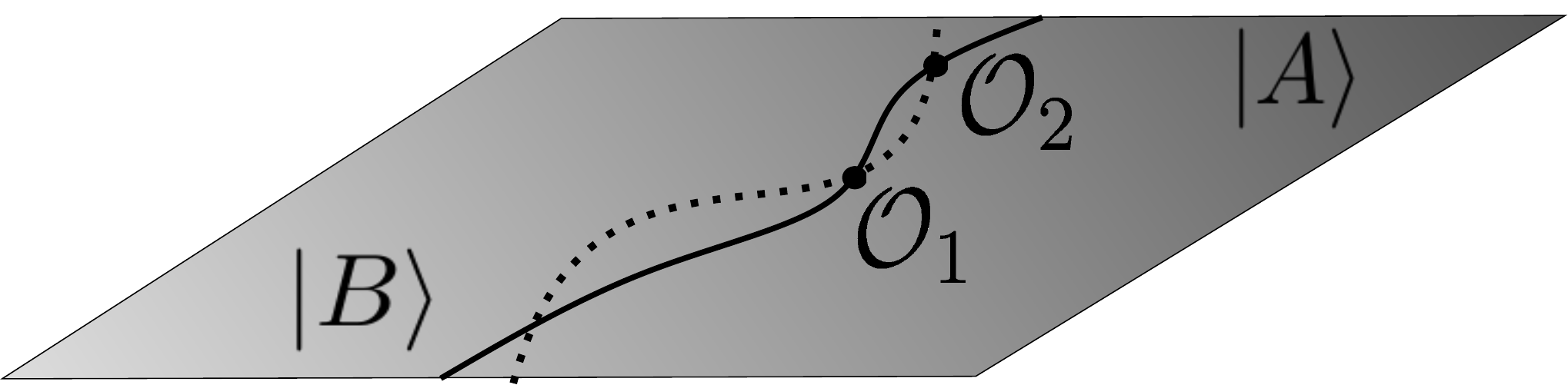}

\caption{\label{fig:Partitioning-a-system}Partitioning a system using an arbitrary
cut establishes a fictitious boundary. The original ground state is
re-written using the basis of each partitions, $\left\{ \left|A\right\rangle ,\left|B\right\rangle \right\} $.
The expectation value of operators defined at the boundary, $\left\{ {\cal O}_{i}\right\} $,
can be done using only \textquotedbl boundary states\textquotedbl{}
in $\left\{ \left|A\right\rangle ,\left|B\right\rangle \right\} $.
Many partitions are possible and they correspond to different 1D channels
that correlations can propagate in the bulk.}
\end{figure}

The state $\left|g\right\rangle $ can always be expressed using the
Schmidt decomposition $\left|g\right\rangle =\sum_{i}\lambda_{i}\left|A_{i}\right\rangle \left|B_{i}\right\rangle $,
where $\left|A_{i}/B_{i}\right\rangle $ form a complete basis of
each partition. These bases comprise both \emph{bulk} and\emph{ boundary}
states. The key observation is that \emph{bulk} states must vanish
at the boundary. As a consequence, any set of operators, $\left\{ {\cal O}_{i}\right\} $,
that live at the cut defining the partitions will couple only to the
\emph{boundary} states of the basis \citep{qi_general_2012}. When
calculating expectation values of $\left\{ {\cal O}_{i}\right\} $,
the state $\left|g\right\rangle $ can be truncated to include only
the \emph{boundary} states, leading to the definition of two boundary
theories,$H_{A}^{{\cal B}}$ and $H_{B}^{{\cal B}}$, which can be
used to label these states. In this context, these boundary theories
can be called as \emph{entanglement models}. Notably, it has been
demonstrated that if the boundary theory is conformal\citep{qi_general_2012},
the maximally entangled conformal state, known as the Ishibashi state
$\left|g_{*}^{{\cal B}}\right\rangle =\sum_{i}\left|{\cal B}_{A,i}\right\rangle \left|{\cal B}_{B,i}\right\rangle $,
is projected into the desired wave function by the \emph{extrapolation
length} $\beta$,

\begin{equation}
\left|g^{{\cal B}}\right\rangle =e^{-\frac{\beta}{2}\left(H_{A}^{{\cal B}}+H_{B}^{{\cal B}}\right)}\left|g_{*}^{{\cal B}}\right\rangle ,\label{eq:edge_theory}
\end{equation}
where $\left|g^{{\cal B}}\right\rangle $ is the truncated ground
state. A similar discussion can also be found in \citep{chen_quantum_2013,swingle_geometric_2012}
and it was established that $\beta^{-1}=T_{E}$ is proportional to
the bulk gap of the topological insulator.

A direct consequence of Eq. (\ref{eq:edge_theory}) is that any two
operators situated on the fictitious boundary will have exponentially
decaying correlations. This observation can be understood within the
framework of the Lieb-Robinson bound\citep{bravyi_lieb-robinson_2006,hastings_locality_2010,sims_lieb-robinson_2010}.
In its more restricted form, the bound says that for a system with
a uniform gap and local interactions, the two point correlation function
of local operators ${\cal O}_{1}$ and ${\cal O}_{2}$ calculated
over the ground state should decay exponentially. Thus, the fictitious
temperature on Eq. (\ref{eq:edge_theory}) enforces the Lieb-Robinson
bound along the direction of the cut. 

The degree of arbitrariness in choosing the cut that creates the partitions
is an important question. When evaluating the expectation value of
an operator at the same spatial position, the particular cut is irrelevant.
The main advantage to use the spectrum of entanglement in this case
is to leverage the knowledge of the boundary theory and its symmetries.
This approach enables us to incorporate, for instance, fermionic interactions
into the analysis, expanding the scope of the discussion.

However, it is when evaluating observables $\left\{ {\cal O}_{i}\right\} $
at different positions we truly capture the underlying physics (see
Fig. (\ref{fig:Partitioning-a-system})). It is instructive to initially
consider a non-interacting theory. Following Ref.~\citep{peschel_calculation_2003},
it is straightforward that for any tight-binding model the one-particle
propagator $C_{ij}=\left\langle g\right|c_{i}^{\dagger}c_{j}\left|g\right\rangle $
on a lattice ${\cal L}$ can be rewritten as $C_{ij}=\text{tr}\left({\cal K}e^{-{\cal H}}c_{i}^{\dagger}c_{j}\right)$
over a subset of points $\left\{ i,j\right\} \in{\cal M}\subset{\cal L}$,
with the spectral of entanglement

\[
{\cal H}=\sum_{\left\{ n,m\right\} \subset{\cal M}}H_{nm}c_{n}^{\dagger}c_{m}
\]
and ${\cal K}$ a normalization constant. A unitary transformation
$U$, that is ${\cal M}$ dependent, diagonalizes the entanglement
spectrum and separates the bulk and boundary modes, i.e.,$U^{\dagger}{\cal H}U={\cal H}_{\text{boundary}}+{\cal H_{\text{bulk}}}$.
In particular, if we choose $\left\{ i,j\right\} $ on the boundary
of ${\cal M}$ the projection of $c_{i}^{\dagger}c_{j}$ over the
boundary modes guarantees that we acquire the same original propagator
regardless of ${\cal M}.$ Consequently, we possess the liberty to
opt for the partition that yields the simplest mode expansion. Generally,
this corresponds to a partition characterized by reflection symmetry,
consequently implying a boundary theory with the shortest distance
between points $i$ and $j$. This argument holds true for an interacting
theory as well since the choice of ${\cal M}$ does not change the
correlation function that is being evaluated. We now use the Kane-Mele
model to illustrate the use of the spectrum of entanglement as a calculation
tool.

\emph{Kane-Mele model with an IRLM:} Arguably the simplest topological
model that preserves time reversal symmetry is the Kane-Mele model
of free fermions on a honeycomb lattice\citep{kane_quantum_2005}.
Let us consider this model in its standard form\citep{kane_quantum_2005}
and introduce a Hubbard interaction between the fermions 

\begin{align}
H_{0} & =\sum_{\left\langle i,j\right\rangle ,\sigma}c_{i\sigma}^{\dagger}c_{j\sigma}+i\Delta\sum_{\left\langle \left\langle i,j\right\rangle \right\rangle ,\sigma,\sigma^{\prime}}\nu_{ij}s_{\sigma\sigma^{\prime}}^{z}c_{i\sigma}^{\dagger}c_{j\sigma^{\prime}}\nonumber \\
 & +u\sum_{i}n_{i\uparrow}n_{i\downarrow}\label{eq:kane-mele}
\end{align}
where $n_{i\sigma}=c_{i\sigma}^{\dagger}c_{i\sigma}$, $\Delta$ in
the next nearest neighbor hopping strength, $\nu_{ij}=\pm1$ depending
on the hopping orientation, $s^{z}$ in the $z$ Pauli matrix and
$u$ is the Hubbard interaction strength. 

\begin{figure}
\begin{centering}
\includegraphics[width=0.6\columnwidth]{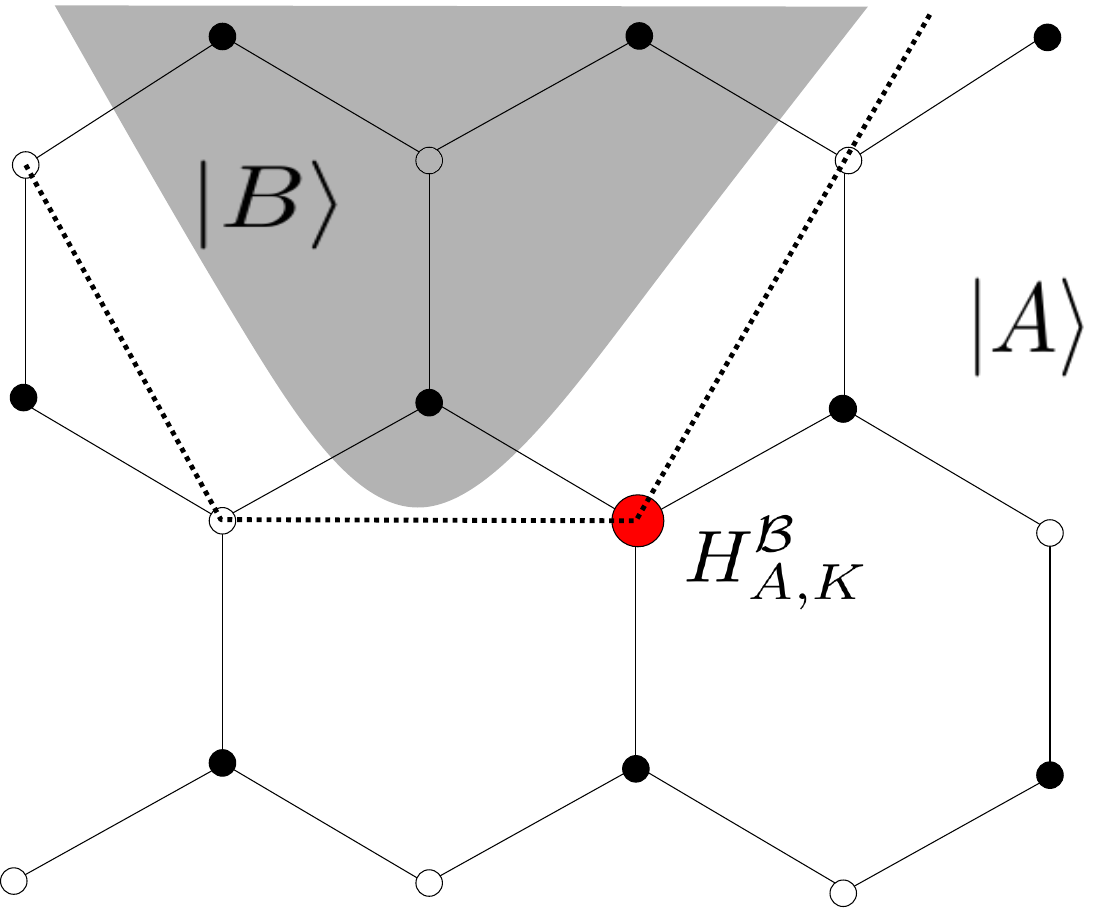}
\par\end{centering}
\caption{\label{fig:Kane-Mele-1}(color on-line) The Kane-Mele model with Coulomb
interactions between the electrons. A magnetic impurity is represented
as a red dot. A fictitious cut creates the two partitions, $\left\{ \left|A\right\rangle ,\left|B\right\rangle \right\} $,
and zip-zag edges. The boundary theory on partition $\left|A\right\rangle $
is $H_{A,K}^{{\cal B}}$, a Kondo model in a L.L. at finite temperature
$T_{E}$, represented by the dotted line.}
\end{figure}

In Fig (\ref{fig:Kane-Mele-1}), we introduce a cut that separates
the two sublattices of the Kane-Mele model, resulting in two zig-zag
edges that can be described in terms of Luttinger liquids\citep{kane_quantum_2005}.
These edges exhibit fermionic currents, namely $J_{A/B}^{R\uparrow}$
and $J_{A/B}^{L\downarrow}$. The fact that the model preserves time-reserval
symmetry implies that at low energies, only marginal forward-scattering
interactions need to be considered. These interactions can be described
by the boundary Hamiltonians
\begin{equation}
H_{A/B,I}^{{\cal B}}=\frac{v_{f}}{4\pi}\int dx\sum_{a\neq b\in\left\{ R\uparrow,L\downarrow\right\} }\left[g_{2}J_{A/B}^{a}J_{A/B}^{b}+g_{4}\left(J_{A/B}^{a}\right)^{2}\right],
\end{equation}
where $v_{f}$ represents the Fermi velocity. By employing Abelian
bozonization\citep{von_delft_bosonization_1998}, the low energy theory
of \emph{each} boundary shown in Figure \ref{fig:Kane-Mele-1}) can
be expressed as

\begin{equation}
H_{A/B,0}^{{\cal B}}+H_{I}^{{\cal B}}=\frac{\tilde{v}}{8\pi}\int_{-\infty}^{\infty}dx\frac{1}{g}\partial_{x}\phi_{A/B}^{2}+g\partial_{x}\theta_{A/B}^{2},
\end{equation}
where $\left[\phi_{A/B}\left(a\right),\partial_{x}\theta_{A/B}\left(b\right)\right]=-4\pi i\delta\left(a-b\right)$,
$g=\sqrt{\left(1+g_{4}-g_{2}\right)/\left(1+g_{4}+g_{2}\right)}$
is the Luttinger parameter and $\tilde{v}=v_{f}\sqrt{\left(1+g_{4}-g_{2}\right)\left(1+g_{4}+g_{2}\right)}$
is the bosonic velocity. The energy gap and the interactions in the
bulk system determine the Fermi velocity and the band width, which
can be approximated as $\Lambda\sim T_{E}$.

At a site on the edge of partition $A$ we introduce a single-impurity
Anderson model with on-site repulsive interactions $H_{A,1}^{{\cal B}}=H_{A,0}^{{\cal B}}+V\sum_{\sigma}c_{0,\sigma}^{\dagger}d_{\sigma}+h.c.+UN_{\uparrow}N_{\downarrow},$
with $N_{\sigma}=d_{\sigma}^{\dagger}d_{\sigma}$. Using the Schrieffer-Wolff
transformation\citep{hewson_kondo_1993} and taking the large $U$
limit, we obtain the Kondo model in a L.L. as \emph{one} of the boundary
theories\citep{lee_kondo_1992,furusaki_kondo_1994,gogolin_bosonization_2004} 

\begin{align}
H_{A,K}^{{\cal B}} & =H_{A,0}^{{\cal B}}+H_{A,I}^{{\cal B}}+J_{z}\partial_{x}\theta_{A}\left(0\right).S^{z}\nonumber \\
 & +\left(J_{\perp}\Lambda\right)e^{-i\phi_{A}\left(0\right)}S^{-}+h.c.\label{eq:kondo_model}
\end{align}
where $J_{k=z,\perp}\sim V^{2}/U$ and $\vec{S}$ the spin projection
of the fermion at the the $d$ level. A single single impurity couple
to the boundary theory is similar to the introduction of a Unruh-DeWitt
detector\citep{crispino_unruh_2008} in the Unhur effect. Therefore,
it completes the analogy introduced by Swingle and Senthil\citep{swingle_geometric_2012}
between the spectrum of entanglement and the Unhur effect. The scaling
equations for this Kondo problem flow to strong coupling for repulsive
interactions, $g<1$, and/or antiferromagnetic Kondo coupling, $J_{z,\perp}>0$.
The Kondo temperature can be estimated using the usual argument as 

\begin{equation}
\frac{T_{K}}{T_{E}}=J_{\perp}{}^{\frac{-1}{1-g+\frac{4J_{z}}{\tilde{v}}}}.\label{eq:kondo_temp}
\end{equation}
The physics of the ground wave function will depend on a comparison
between the Kondo temperature, $T_{k}$, and the temperature of entanglement,
$T_{E}$, which acts as an infrared cut-off to the renormalization
process. 

\begin{figure}
\begin{minipage}[t]{0.45\columnwidth}%
\begin{center}
\includegraphics[width=0.9\columnwidth]{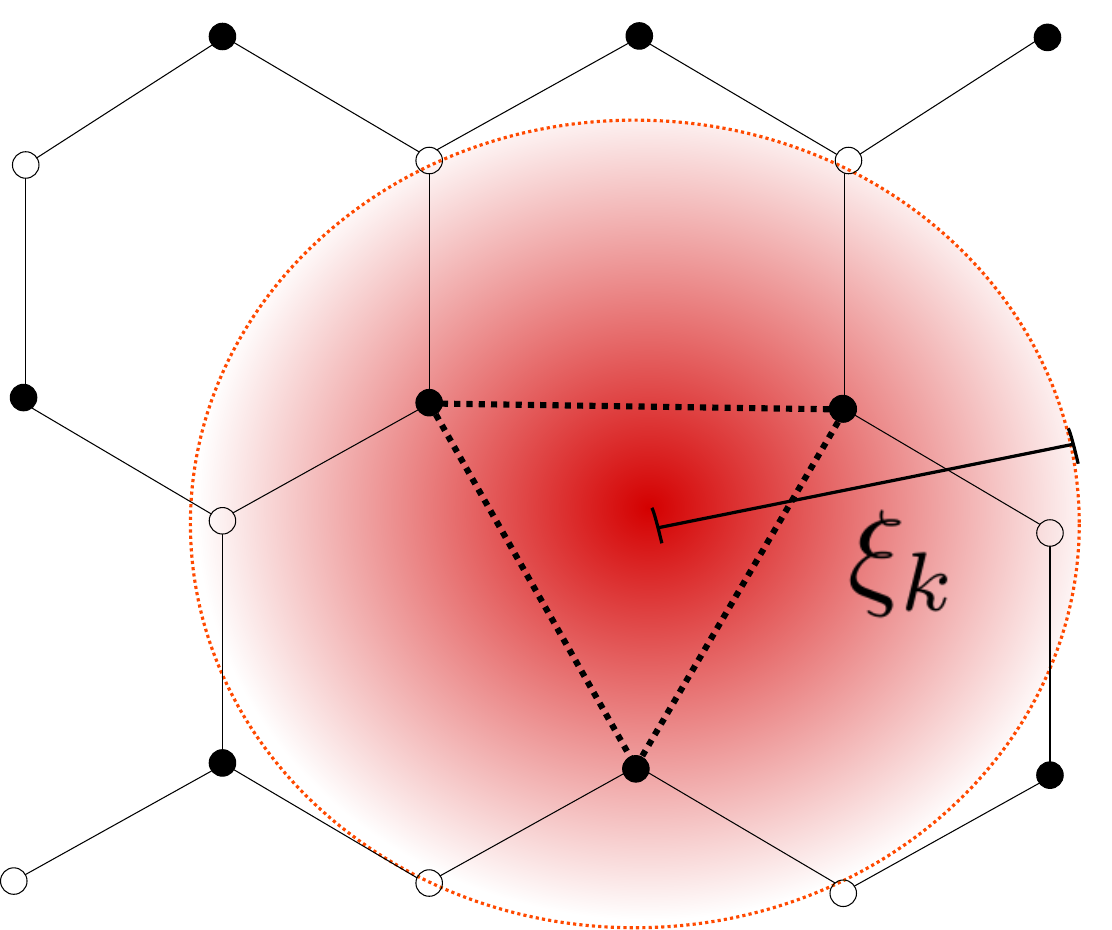}
\par\end{center}
\begin{center}
(a)
\par\end{center}%
\end{minipage}%
\begin{minipage}[t]{0.45\columnwidth}%
\begin{center}
\includegraphics[width=0.9\columnwidth]{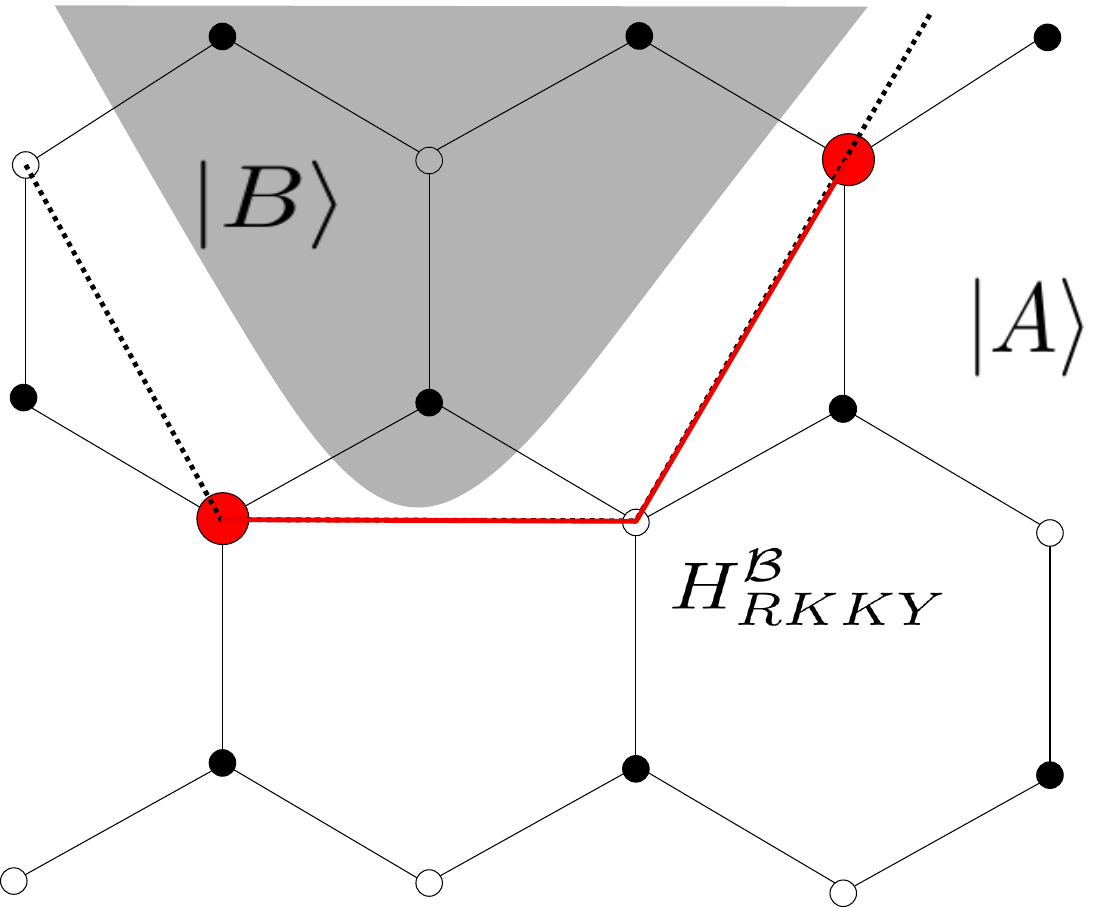}
\par\end{center}
\begin{center}
(b)
\par\end{center}%
\end{minipage}

\caption{\label{fig:Kane-Mele-2}(color on-line) (a) If $T_{E}\ll T_{k}$ a
Kondo singlet is formed (red region of length $\xi_{k}$). We expect
that edge states exist at the end of the Kondo cloud (depicted as
dotted lines in the figure). (b) In the perturbative regime, $T_{k}\lesssim T_{E}$,
an effective $RKKY$ interaction controls the dynamics of the magnetic
impurities.}
\end{figure}

For repulsive interactions and when $T_{k}\gg T_{E}$ the Kondo singlet
forms. In the limiting case of $J_{z,\perp}\to\infty$, a $c$-fermion
at the the ressonat level site is bound to form the Kondo singlet.
This effectively removes the site from the lattice and introduces
a zig-zag edge in the middle of the bulk, see Fig. \ref{fig:Kane-Mele-2}a.
However, for a finite $J_{z,\perp}$ the many body state is more complex.
The fermion at the local level $d$ is entangled with many electrons
on the $c$-band, forming what is known as the Kondo cloud\citep{v_borzenets_observation_2020}.
The Kondo cloud is not a well define entity in real space, but it
is usually assumed to have a characteristic length scale of $\xi_{k}=\tilde{v}/T_{k}$.
This region of the material, characterized by the Kondo cloud, is
an intriguing state of matter, that should be surrounded by gapless
real edge states. If a diluted set of these Kondo impurities is separated
by distances of the order of $\xi_{k}$, it could allow current to
pass through the bulk of the material. This implies that for a linear
dimension $L$, a critical density of magnetic impurities of $n_{c}=\xi_{k}/L$
would melt the topological phase. The tunneling of electrons in internal
edge states created by impurities have been proposed before as a mechanism
to destabilize the topological phase\citep{shen_topological_2017}.
The difference here is that the region is not related to size of the
impurities, but rather to the length of the Kondo cloud. 

In the limit where $T_{k}\lesssim T_{E}$, we enter a perturbative
regime where the renormalized $J_{z,\perp}$ remains small due to
an irrelevant flow or an infrared cutoff. In this regime, multiple
magnetic impurities can interact through the $c$-band. To study this
interaction, we need to evaluate the Green's functions of the $c$-electrons
that mediate the interaction. For simplicity, let us consider a pair
of magnetic impurities, $\left\{ \vec{S}_{1},\vec{S}_{2}\right\} $,
located on the same sublattice of the honeycomb lattice, see Fig.
\ref{fig:Kane-Mele-2}b. We introduce a partition that connects the
two impurities through a minimum path of distance $\Delta x$ along
the edge. Since we are in the perturbative regime, it is straightforward
to write a Ruderman-Kittel-Kasuya-Yosida (RRKY) like interaction between
the two impurities using the finite temperature Green's function of
the L.L.\citep{von_delft_bosonization_1998},

\begin{equation}
H_{RKKY}^{{\cal B}}\propto\frac{J_{k}^{2}}{\left[\frac{\beta_{E}}{\pi}\sinh\left(\frac{\pi}{\beta_{E}}\Delta x\right)\right]^{2g}}\vec{S}_{1}.\vec{S}_{2}+\text{l.r.t.}\,.
\end{equation}
The ground state of several pairs of these magnetic impurities are
singlets. Therefore, by measuring the distribution of biding energies
of these singlets, and independently measuring the bulk gap, it is
possible to directly determine the Luttinger parameter.

The physics of a dense array of magnetic impurities will ultimately
depend on the interplay of the three energy scales in the problem:
$T_{E}$, $T_{K}$, $T_{RRKY}\sim J_{k}^{2}\tilde{v}$. This interplay
opens up a wide range of possibilities for exploring the rich phenomenology
of heavy fermions.

\emph{Conclusions:} In this manuscript we propose a new theoretical
approach for incorporating electronic interactions in the study of
two dimensional topological insulators. Building upon the established
correspondence between edge state physics and the entanglement spectrum
in gapped topological systems, we leverage this connection to deconstruct
the two dimensional system as one dimensional channels that connect
any two points within in the system. The effective theory of these
channels is precisely the edge state theory at the fictitious temperature
of entanglement. The effective theory carries the same topological
protection that real edge states have. Consequently, we can elegantly
incorporate fermionic interactions into the framework. To analyze
these systems we can employ well-known techniques such as Abelian
bosonazation or te density matrix renormalization group. While considering
one dimensional channels in topological insulators is not a new idea\citep{kane_fractional_2002,teo_luttinger_2014,tam_nondiagonal_2021,neupert_wire_2014,iadecola_wire_2016},
our approach naturally emerges from the ground state wave function.

We follow the general discussion by considering the Kane-Mele model
with interactions and in the presence of magnetic impurities. It is
well-established that impurities can give rise to localized regions,
or islands, within a topological system, which harbor edge states\citep{shen_topological_2017}.
Our findings reveal two intriguing scenarios. Firstly, in the Kondo
regime we identify a possible mechanism to destabilize the topological
phase. Secondly, in the perturbative regime we observe that the impurities
can order magnetically through a RRKY-like interaction. Interestingly,
this magnetic ordering in a dilute impurity system presents a unique
opportunity to experimentally measure the strength of fermionic interactions
in the topological system.
\begin{acknowledgments}
M. Derli would like to thank CAPES for financial support. E. Novais
would like to thank useful discussions with T. C. Jordão, F. M. Hungria,
A. Landulfo and E. Miranda.
\end{acknowledgments}

\bibliographystyle{apsrev}

\end{document}